\documentclass[a4paper,fleqn,usenatbib]{mnras}

\usepackage{newtxtext,newtxmath}

\usepackage{caption}

\usepackage[T1]{fontenc}
\usepackage{ae,aecompl}

\usepackage{threeparttable, tablefootnote} 

\usepackage{graphicx}	
\usepackage{amsmath}	
\usepackage{amssymb}	

\title[Radial evolution of pure high-speed streams]{Radial evolution of the solar wind in pure high-speed streams: HELIOS revised observations}

\author[D. Perrone et al.]{
Denise Perrone,$^{1}$\thanks{E-mail: d.perrone@imperial.ac.uk}
D. Stansby,$^{1}$
T. Horbury$^{1}$
and L. Matteini$^{2}$
\\
$^{1}$Department of Physics, Imperial College London, London SW7 2AZ, UK\\
$^{2}$LESIA, Observatoire de Paris, PSL Research University, CNRS, Sorbonne University, UPMC Univ. Paris 06, Univ. Paris Diderot, Sorbonne Paris Cité, France 
}

\date{Accepted XXX. Received YYY; in original form ZZZ}

\pubyear{2018}

\begin{document}
\label{firstpage}
\pagerange{\pageref{firstpage}--\pageref{lastpage}}
\maketitle

\begin{abstract}
Spacecraft observations have shown that the proton temperature in the solar wind falls off with radial distance more slowly than expected for an adiabatic prediction. Usually, previous studies have been focused on the evolution of the solar-wind plasma by using the bulk speed as an order parameter to discriminate different regimes. 
In contrast, here, we study the radial evolution of pure and homogeneous fast streams (i.e. well-defined streams of coronal-hole plasma that maintain their identity during several solar rotations) by means of re-processed particle data, from the HELIOS satellites between $0.3$ and $1$~AU. We have identified 16 intervals of unperturbed high-speed coronal hole plasma, from three different sources and measured at different radial distances. The observations show that, for all three streams, (i) the proton density decreases as expected for a radially expanding plasma, unlike previous analysis that found a slower decrease; (ii) the magnetic field deviates from the Parker prediction, with the radial and tangential components decreasing more slowly and quickly than expected, respectively; (iii) the double-adiabatic invariants are violated and an increase of entropy is observed; (iv) the proton-core temperature anisotropy is constrained by mirror mode instability; (v) the collisional frequency is not constant, but decreases as the plasma travels away from the Sun. 
The present work provides an insight into the heating problem in pure fast solar wind, fitting in the context of the next solar missions, and, especially for Parker Solar Probe, it enables us to predict the high-speed solar-wind environment much closer to the Sun.
\end{abstract}




\section{Introduction}

Since the prediction and discovery of the solar wind, the puzzle of ion acceleration and heating has become one of the most compelling problems in space plasma physics. However, many basic ideas have never been rigorously confronted by direct measurements in the region where these processes are actually occurring. The only possibility to obtain an answer to these questions is through in situ measurements of the solar-wind plasma as close as possible to the Sun, being the main goal of the next solar missions, i.e. NASA's Parker Solar Probe~\citep{fox16} and ESA's Solar Orbiter~\citep{mul13}.  

To date, the only mission specifically designed to explore the interplanetary medium in the near-solar environment were the HELIOS solar probes. Two almost identical spacecraft were launched into highly elliptical and small inclination orbits, with a perihelion distance of 0.31~AU for HELIOS1 (launched on December 10$^{th}$, 1974) and 0.29~AU for HELIOS2 (launched on January 15$^{th}$, 1976)~\citep{sch90}. 
One of the main goals of the mission was to provide, for the first time, full three-dimensional ion velocity distributions. These displayed strong departures from the thermodynamic equilibrium, with different parallel and perpendicular temperatures with respect to the background magnetic field. In addition, proton distribution functions often showed a field-aligned beam, whose relative velocity with respect to the core (which contains $\sim 80\%$ of the proton population) is of the order of the local Alfv\'en speed~\citep{mar82b}. The complex shapes of the ion velocity distributions, typically far from Maxwellian distributions, confirm that Coulomb collisions are not able to keep the solar-wind system in thermodynamic equilibrium. However, the role of collisions depends on the properties of the solar wind and varies with distance and latitude~\citep{mat11}. In particular, collisions influence the ion velocity distributions in slow solar wind~\citep{liv86}, working to remove ion anisotropies and the relative proton-alpha drift, while their role in fast solar wind can be considered negligible~\citep{margol83}.   

Another important goal of the HELIOS mission was to give access to the heliocentric variation of the solar wind properties in the inner heliosphere. Particularly, the studies on the thermodynamics of the protons have highlighted the need for an efficient proton heating in order to explain the observations. In fact, HELIOS measurements have shown that the proton temperature falls off more slowly than expected for an adiabatic prediction. 
For a radially expanding and collisional plasma that cools adiabatically with a polytropic index of $5/3$, the temperature variation with heliocentric distance is expected to be proportional to $R^{-4/3}$. In the case of a collisionless plasma, instead, where different parallel and perpendicular temperatures are present, as in the case of the solar wind, the variations of $T_{\parallel}$ and $T_{\perp}$ should be described (neglecting collisions and heat fluxes) by the double-adiabatic hypothesis (or CGL)~\citep{cgl56} based on the two adiabatic invariants $T_{\perp}/B$ and $T_{\parallel}B^2/n^2$ (i.e. $T_{\perp} \propto R^{-2}$ and $T_{\parallel} \sim$~const for a spherical expansion, in presence of a strictly radial magnetic field and a constant solar-wind velocity). However, in the solar wind, these invariants are observed to be broken~\citep{mar83}, due to the action of turbulent dissipation or wave damping~\citep{mar87}. In fact, in situ measurements show that $T_{\perp}$ decreases in the expansion more slowly, whereas $T_{\parallel}$ decreases faster, than expected~\citep{mar82b,hel11,hel13}. Some of the observed properties are compatible with the interaction of the ions with high-frequency Alfv\'en cyclotron waves~\citep[e.g.][]{hol02}, and signatures of cyclotron heating with the formation of quasilinear plateaux are also observed~\citep{heu07}. It is worth noting that the analyses cited above do not separately consider the proton core and beam. In particular, \citet{hel11,hel13} used numerical moments of the proton distribution function instead of analytical fits, meaning that the contribution of the proton core and beam populations cannot be separated; while in \citet{mar82b}, although the fit of the proton velocity distribution isolates the two proton populations, the radial trend for the parallel and perpendicular temperatures is given by considering the whole velocity distribution. 

Over the years, several studies have been focused on the radial evolution of the plasma properties in different regimes of solar wind by using the bulk speed as an order parameter, i.e. by using a threshold for the proton radial velocity~\citep{hel11,hel13} or averages over 100~km/s wide speed intervals~\citep[e.g.][]{mar82b}. However, both these choices can be limited and approximate, since streams with different origins or from heterogeneous solar wind types could be taken into account, potentially leading to erroneous conclusions. In this respect, an ideal approach, in order to remove any additional effects, is to study the characteristics of the solar-wind radial evolution by using a well-defined stream that maintains its identity during the radial expansion. A high-velocity stream, observed by HELIOS spacecraft during three successive solar rotations and at different distances from the Sun, has been used to study the radial evolution of the power spectra of the MHD turbulence by means of magnetic field data~\citep{bav81,bav82a,bav82b}. These intervals have been selected by visual inspection of hourly averages of the solar wind data and the radial dependence for the temperature is found to be $\propto R^{-0.9}$~\citep{ver95}. 

High-speed streams are always present in the heliosphere. Around a solar minimum, such fast wind originates in large and stable long-lived polar, and occasionally low-latitude, coronal holes and recurs at intervals of $\sim 27$ days; while, near a maximum of the solar activity, when the polar holes disappear, shorter lived high-speed wind is issued generally from more transient coronal holes located in active regions~\citep{hun72,fou02}. Moving radially outwards, they interact with the preceding slower solar wind, forming a region of compressed plasma along the leading edge of the stream. The stream interface, i.e. the boundary between slow and fast wind, is typically characterised by a sharp density drop, temperature rise and specific entropy increase~\citep{ric18}.     

In the present work, we study the radial evolution of pure and homogeneous high-speed streams in the inner heliosphere. In order to avoid any additional effects due to the presence of interaction regions (e.g. acceleration/deceleration of the streams), we select only intervals of unperturbed coronal-hole plasma, from different sources, and we follow their radial evolution during several solar rotations. Independently of the source, we find that the proton density decreases as expected for a radially expanding plasma, i.e. $\propto R^{-2}$, unlike the previous studies that have shown a slower decrease. 
Moreover, we observe heating in the proton core population, with a violation of the double-adiabatic invariants and an increase of entropy. Furthermore, the proton core temperature anisotropy appears to be constrained by the mirror mode instability. Finally, we recover that the magnetic field deviates from the Parker prediction, with the radial component that decreases more slowly and the tangential component that decreases more quickly than expected.

\section{HIGH-SPEED PLASMA STREAMS}

\begin{figure}
	\includegraphics[width=\columnwidth]{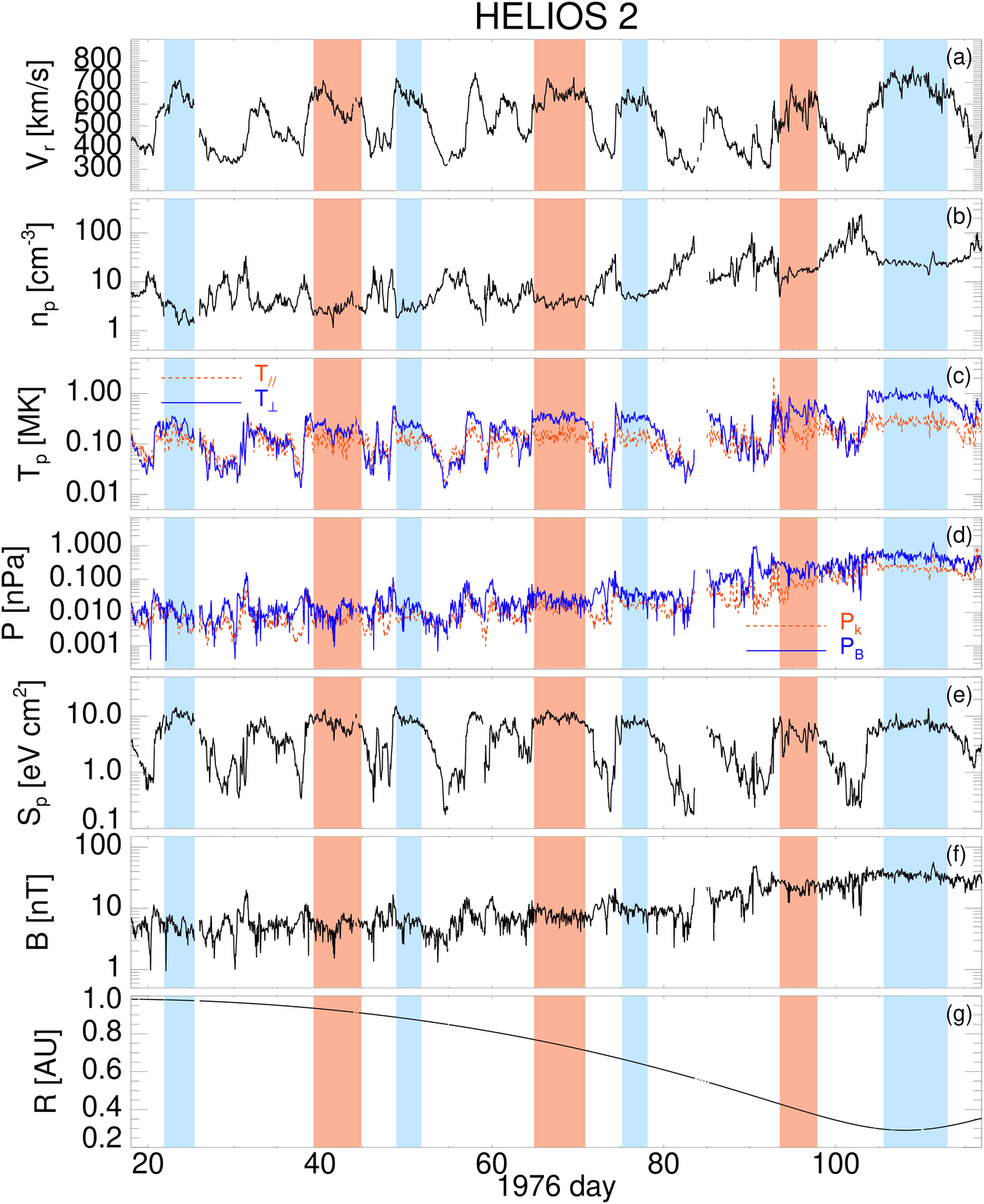}
    \caption{Overview of solar-wind data, hourly averaged, during the primary mission of HELIOS2 
    			(January to April 1976).
    			From top to bottom: (a) radial component of the velocity; (b) proton density; (c) proton 
			temperature: $T_{\parallel}$ in red-dashed line and $T_{\perp}$ in blue-solid line; 
			(d) proton kinetic pressure, $P_{k}$, in 
			red-dashed line and magnetic pressure, $P_B$, in blue-solid line; 
			(e) proton specific entropy; (f) magnitude of the magnetic field; and (g) radial distance. 
			Color-filled bands denote unperturbed coronal-hole plasma streams. 
			Different colours indicate recurrent streams of different origin.}
    \label{fig:stream}
\end{figure}

We use data from the twin HELIOS probes in the first two years of their mission, which correspond to a declining phase of solar activity when the interplanetary structure was characterised by stable recurrent high-speed streams. The study is performed on a new data set~\citep{sta17,sta18} which contains estimates of number density, velocity and temperatures of the proton core population in the solar wind, obtained from a systematic fitting with bi-Maxwellian functions of all the original HELIOS 3D distribution functions. Magnetic field data are also provided as an average from the values taken whilst the distribution function was measured. 

An overview of the primary mission of HELIOS2 (January to April 1976) is summarised in Fig.~\ref{fig:stream}, where hourly averaged data are used. Panel (a) shows the radial component of the solar wind speed, with the characteristic structure of recurrent rapid rises (over about 1 day) and slower decays of the bulk speed. A similar pattern is present for the proton parallel ($T_{\parallel}$, red-dashed line), perpendicular ($T_{\perp}$, blue-solid line) and total ($T_p=(T_{\parallel}+2T_{\perp})/3$, not shown) temperatures in panel (c); and for the proton specific entropy, $S_p=T_p/n_p^{2/3}$, in panel (e). The profile of $S_p$ clearly shows a sudden jump where the spacecraft crossed regions from low-entropy streamer-belt plasma into high-entropy coronal-hole plasma. 
The proton density $n_p$ (panel b) and magnetic intensity $B$ (panel f) were enhanced near the leading edges of the streams, due to the dynamical interaction of the streams with their surroundings. A similar behaviour is observed in panel (d) for both the proton kinetic ($P_{k}=n_pk_BT_p$, red-dashed line) and magnetic ($P_{B}=B^2/8\pi$, blue-solid line) pressures.

Finally, color filled bands in Fig.~\ref{fig:stream} denote unperturbed coronal-hole plasma streams identified using the procedure described in Section~\ref{sec:iden}. 
The two colours of the filled bands indicate two recurrent ($\sim 27$~days) streams from different sources, observed at different radial distances (panel g) over three solar rotations.

\subsection{Unperturbed coronal hole plasma}
\label{sec:iden}

\begin{figure}
	\includegraphics[width=\columnwidth]{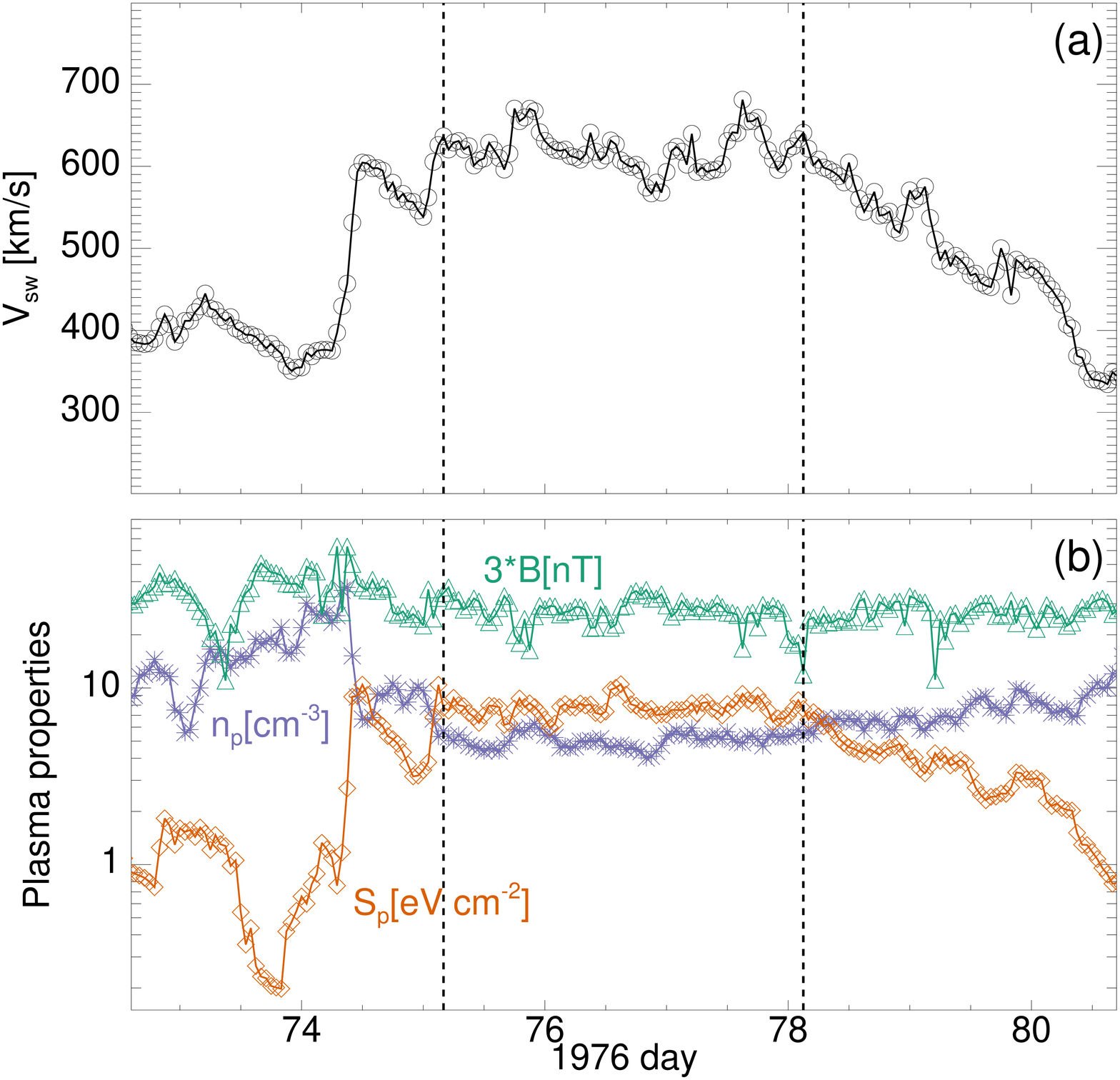}
    \caption{Example of unperturbed coronal-hole plasma. Hourly averaged data of  
    		solar-wind speed in panel (a), and proton density (violet-star line), magnetic 
		field (green-triangle line) and proton specific entropy (orange-diamond line) in 
		panel (b). Vertical black-dashed lines indicate onset and end of the 
		flattop, respectively.}
    \label{fig:selection}
\end{figure}

In general, fast solar wind from coronal-hole regions is characterised by low number densities, high proton entropies and low heavy-ion charge-state ratios~\citep{gei95}. Moreover, protons in fast solar wind exhibit heating and an increase of the specific entropy with radial distance in the inner heliosphere~\citep{mar82b,mar83,hel11,bor14}. Here, rather than taking a cutoff in speed, we select more restricted periods within high-speed streams, characterised by a `flat-top-like' shape in the temporal profile of the solar wind velocity and approximately constant values for the magnetic field strength, proton density and specific entropy~\citep{bor16}. This choice allows us to avoid the compressed coronal-hole plasma, at the end of the corotating interaction regions, and the rarefaction regions, on the trailing edge of the high-speed streams. An example of the identification procedure is shown in Fig.~\ref{fig:selection}. 

Following~\citet{bor16}, the onset of the unperturbed coronal-hole interval is chosen when the magnetic field (green-triangle line) and the proton density (violet-star line) decline to a more or less steady values early in the flattop region. Moreover, at the same time, the proton specific entropy (orange-diamond line) reaches a maximum and steady value. The end of the interval, which corresponds to the onset of the rarefaction region, is taken when the solar wind velocity starts to systematically decrease and also the proton specific entropy starts to decline from high (and steady) values towards lower values. The same behaviour of $S_p$ is also observed for the proton temperature (not shown here).

\begin{table}
	\centering
	\caption{Intervals of unperturbed coronal-hole plasma in both 
	HELIOS1 and HELIOS2 used in this study.}
	\label{tab:intervals}
	\begin{tabular}{cccccccc}
		\hline
		\cline{3-4} \cline{5-6}
                      &   & \multicolumn{2}{ c }{Start} & \multicolumn{2}{ c }{End} \\
		\cline{3-4} \cline{5-6}
		\hline
		Spacecraft & Year & Day & UT & Day & UT & $R$ (AU) & Flattop \\
		\hline	
		HELIOS1 & 1975 & 13 & 04 & 17 & 02     & 0.88 & C1 \\
		HELIOS1 & 1975 & 40 & 08 & 45 & 21     & 0.64 & C2 \\
		HELIOS1 & 1975 & 72 & 16 & 75 & 02     & 0.31 & C3 \\
		HELIOS1 & 1975 & 306 & 08 & 308 & 00 & 0.76 & A1 \\
		HELIOS1 & 1975 & 359 & 12 & 361 & 10 & 0.99 & A2 \\
		HELIOS2 & 1976 & 21 & 21 & 25 & 10     & 0.98 & A3 \\
		HELIOS2 & 1976 & 39 & 06 & 44 & 20     & 0.92 & B1 \\
		HELIOS1 & 1976 & 47 & 02 & 49 & 14     & 0.74 & A4 \\
		HELIOS2 & 1976 & 48 & 21 & 51 & 20     & 0.88 & A5 \\
		HELIOS2 & 1976 & 64 & 22 & 70 & 21     & 0.74 & B2 \\
		HELIOS1 & 1976 & 65 & 20 & 70 & 11     & 0.53 & B3 \\
		HELIOS1 & 1976 & 74 & 10 & 79 & 13     & 0.42 & A6 \\
		HELIOS2 & 1976 & 75 & 04 & 78 & 03     & 0.65 & A7 \\
		HELIOS2 & 1976 & 94 & 16 & 97 & 21     & 0.40 & B4 \\
		HELIOS2 & 1976 & 105 & 14 & 113 & 01 & 0.30 & A8 \\
		HELIOS1 & 1976 & 113 & 10 & 116 & 21 & 0.57 & A9 \\ 
		\hline		
	\end{tabular}
\end{table}

Using the hourly averaged data from both HELIOS1 and HELIOS2 in the solar minimum period between January 1975 and April 1976, we identify 16 intervals of unperturbed coronal-hole plasma from three different coronal holes at different radial distances, listed in Table~\ref{tab:intervals}. Recurrent streams of different origin are indicated with a different letter, while the number indicates the chronological order of the observed streams. 

The choice of recurrent flattops, originating in the same coronal hole, allows the investigation of the radial dependence of the solar-wind characteristic quantities in a homogeneous data set of pure fast solar wind, assuming that variations with heliographic latitude are absent or not important. 
Therefore, it is essentially possible to study the evolution of plasma from the same region as it evolves with distance. Finally, the presence of three homogeneous intervals, from three different coronal holes, allows us to study the dependence on the source of the high-speed streams.   

\subsection{Solar-wind speed}

\begin{figure}
	\includegraphics[width=\columnwidth]{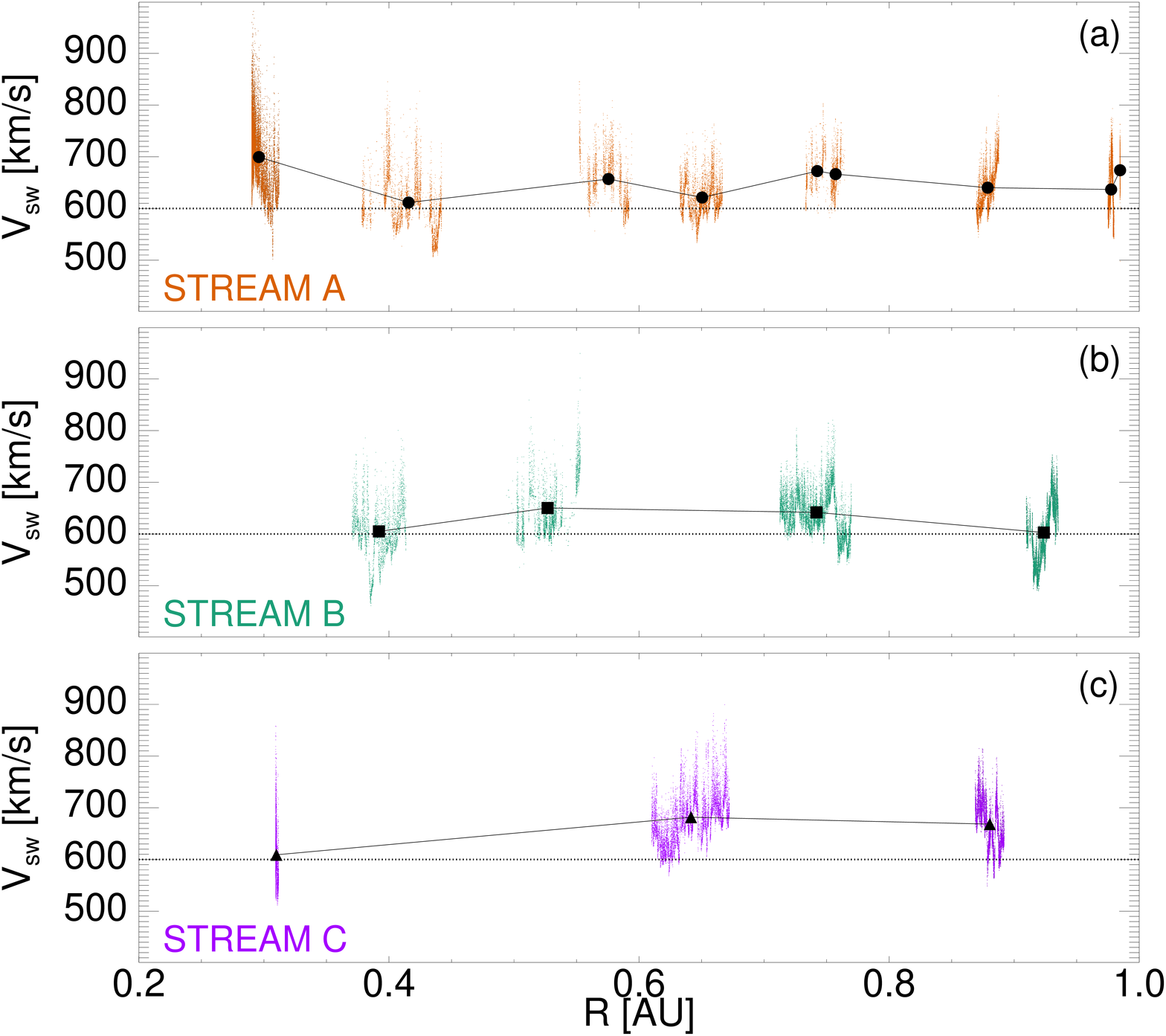}
    \caption{Solar-wind speed, $V_{_{sw}}$, for the three recurrent flattop streams 
    			as a function of the radial distance, $R$. The black symbols indicate the mean of the 
			speed in each interval listed in Table~\ref{tab:intervals}, while the horizontal dotted 
			lines refer to the threshold used by~\citet{hel11} to select fast wind.}
    \label{fig:velocity}
\end{figure}

Fig.~\ref{fig:velocity} displays the evolution of the solar-wind speed, $V_{_{sw}}$, for the three recurrent high-speed streams (different panels for different coronal-hole sources) listed in Table~\ref{tab:intervals}, as a function of the radial distance, $R$. The black symbols in each panel indicate the mean values, for velocity and distance, evaluated within each flattop. Different sources produce streams with different speeds, that can have a wide range of values, sometimes unexpectedly low. Stream B, for example, has flattop intervals with $V_{_{sw}}$ much lower than $600$~km/s, but is still a high-speed stream. This means that fixing a low threshold in velocity for selecting fast solar wind, as in the case discussed in~\citet{hel11}, can give incomplete information. In fact, as observed in Fig.~\ref{fig:velocity}, pure high-speed intervals can have values lower than $600$~km/s (horizontal dotted lines). Moreover, knowing the value of the velocity without the context of the observation produces approximate and eventually erroneous information, since high values of the velocity (i.e. $V_{_{SW}} > 600$~km/s) can be also associated with interaction regions, which have different physics with respect to pure fast solar-wind streams. Finally, Fig.~\ref{fig:velocity} also highlights that the choice to bin the velocity in range of $100$~km/s~\citep[e.g.][]{mar82b} can mix wind from different sources or interaction regions, since the value of the velocity cannot be used alone to discriminate between different types of wind.     

\begin{figure}
	\includegraphics[width=\columnwidth]{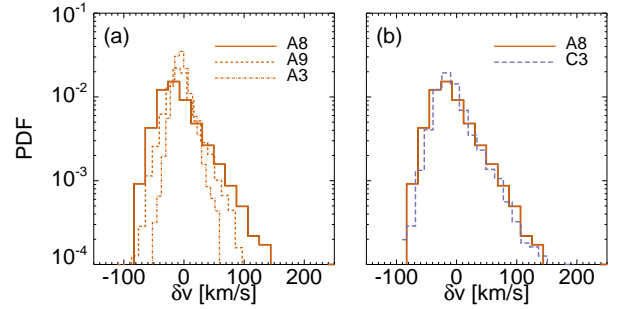}
    \caption{Probability distribution functions of the radial speed with respect to a 30 minute running 
    			mean during (a) three flattops of the same stream at different radial 
			distance, $R \sim 0.3, 0.57, 0.98$~AU, and (b) two flattops of different streams 
			at $\sim 0.3$~AU.}
    \label{fig:spikes}
\end{figure}

Recently, \citet{hor18} have reported the presence of intermittent enhancements in plasma speed in the near-Sun high-speed solar wind, by using 4 days within the flattop A8 listed in Table~\ref{tab:intervals}. These events are Alfv\'enic with an anti-sunward propagation sense in the solar-wind frame and the core proton distribution within them is no different to the background wind~\citep[see also][]{mat14}. However, since the observed spikes reach a speed up to 250~km/s with respect to the ambient wind, they carry a significant fraction of the total momentum and energy of the plasma. Fig.~\ref{fig:spikes} shows the Probability Distribution Functions (PDFs) of the instantaneous radial velocity fluctuations with respect to a 30 minute running average solar-wind speed, i.e. $\delta v=V_r- \langle V_r \rangle$. The solid-orange line, in both panels (a) and (b), shows the PDF of $\delta v$ for the flattop A8, which includes the high-speed stream studied by~\citet{hor18}, characterised by a longer right tail (positively skewed). In panel (a) we compare the PDF of the flattop A8 ($R \sim 0.3$~AU) with the PDF of two other flattops of the same stream but at different radial distances, i.e flattop A9 ($R \sim 0.57$~AU, orange-dashed line) and flattop A3 ($R \sim 0.98$~AU, orange-dash-dotted line). The spikes present lower amplitudes as the stream moves away from the Sun~\citep[see also][]{mat14} and also the skewness of the distributions becomes less important (at 1~AU the distribution is almost symmetric). In panel (b) we show the comparison between the flattop A8 and a flattop from a different stream (i.e. flattop C3, violet-long-dashed line) but at the same distance from the Sun ($R \sim 0.31$~AU). In this case, the PDFs of $\delta v$ are similar, meaning that the importance of the spikes is characterised by the distance from the Sun and is independent of the origin of the specific stream. Moreover, if we plot the PDFs of $\delta v$ normalised to the mean value of the Alfv\'en speed within each interval (not shown here), we found that all the PDFs are comparable. Therefore, the amplitude of the spikes is constrained by the Alfv\'en speed, as expected due to their Alfv\'enic nature~\citep{mat15}.

\section{Radial dependences}

\begin{figure}
	\includegraphics[width=\columnwidth]{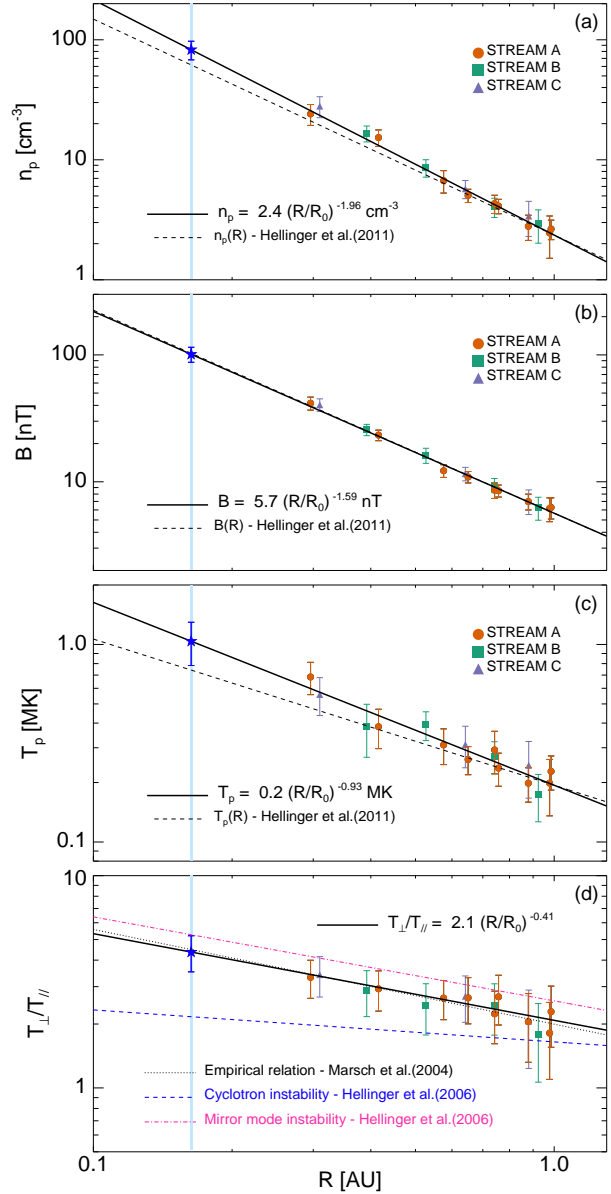}
	\centering
    \caption{Overview of solar-wind radial evolution. From top to bottom: (a) proton density; 
    			(b) magnetic field; (c) proton temperature; and (d) temperature anisotropy. 
			Each point and the relative error bar  
    			refer to the mean and $\pm$ the standard deviation in each flattop. 
			Different colours and symbols indicate different streams.
			Black-solid lines show the fits for the stream A (orange circles).
			Comparisons with previous studies and/or with theoretical predictions 
			are also shown (see legend in each panel). Blue stars in each panel refer to
			extrapolated fast-stream values at $\sim 35 R_S$ (blue-filled band), corresponding to 
			the heliospheric distance of the first three perihelia for Parker Solar Probe.}
    \label{fig:evolution}
\end{figure}

In the following, we will study the radial evolution of several quantities for the three unperturbed coronal-hole high-speed streams listed in Table~\ref{tab:intervals}. In Figs.~\ref{fig:evolution} and~\ref{fig:collisions}, for each flattop, an averaged value of the considered quantity within each interval, $f$, with the relative standard deviation as error, $\pm \sigma_f$, is shown. Different colours and symbols are used for the different streams (see legend). Moreover, we fit the averaged values of stream A by using the least squares linear regression function (although for some of these considered quantities no power law is expected) in logarithmic space, i.e. $\log f=\alpha \log x + k$, with $x=R/R_0$ ($R_0=1$~AU) and $k=\log f_0$, indicated by black-solid lines. The choice of the stream A is due only to statistical reasons, i.e. it is the stream with the most intervals, and this does not influence the generality of the fit results since most of the points of the other streams lie on the same curve. Moreover, comparisons with previous studies and/or with theoretical predictions are shown (see legend in each panel). Finally, expected values in fast streams at $\sim 35 R_S$ (blue-filled band), where $R_S$ is the solar radius, which corresponds to the heliospheric distance of the first three perihelia for Parker Solar Probe, are indicated by blue stars.
 
\subsection{Density}

Panel (a) of Fig.~\ref{fig:evolution} shows the radial dependence of the proton number density, $n_p$, where the fit (black-solid line) gives 
\begin{eqnarray}
n_p=(2.4\pm0.1) (R/R_0)^{-(1.96\pm 0.07)}\ \text{cm}^{-3}
\end{eqnarray} 
Unlike the radial dependence found by~\citet{hel11}, i.e. $n_p \propto R^{-1.8}$, indicated in panel (a) as a black-dashed line (we assume the same value for the densities at $R=1$~AU to better appreciate how they diverge at smaller $R$), we observe that the proton core density decreases as expected for a radially expanding solar wind. In~\citet{hel11}, fast solar wind was chosen for $V_{_{sw}} > 600$~km/s and complete proton distribution functions were used (i.e. a possible contribution to the total density of secondary beams is possible, which increases with the heliocentric distance~\citep{mar82b}). Therefore, a slower decrease for $n_p$ could be due to the presence of both secondary proton beams and interaction regions between fast and slow wind. In our analysis, only the core is considered and compression and/or rarefaction regions are avoided; then, $n_p \propto R^{-2}$. The radial expansion at constant speed of pure fast wind plasma is also confirmed by the analysis of the density flux, $n_p V_r R^2$, which is found to be almost constant (not shown here). In fact, the fit for the stream A gives $n_p V_r \propto R^{-(1.97\pm0.08)}$, implying that no protons are removed from or added to the proton core population during radial expansion.

\subsection{Magnetic field}

The radial dependence of the magnetic field magnitude, $B$, is shown in panel (b) of Fig.~\ref{fig:evolution}, where the fit gives 
\begin{eqnarray}
B=(5.7\pm0.2) (R/R_0)^{-(1.59\pm 0.06)}\ \text{nT}
\end{eqnarray} 
The same dependence, $B \propto R^{-1.6}$, was found in~\citet{hel11}. 

The generally accepted structure of the interplanetary magnetic field refers to Parker's model~\citep{par58,par63}. It is based on a radial expansion of the solar wind at a constant speed, that determines, together with the rotation of the Sun, the configuration of the magnetic field. 
According to the model, the radial component of the magnetic field should decrease as $R^{-2}$, while the tangential component should vary as $R^{-1}$. In this context, these expected radial evolutions are often used to make extrapolations about the magnetic field from one heliocentric distance to another as, for example, in space weather models~\citep[e.g.][]{ril14}. 

Here, in a minimum phase of solar activity (i.e. the coronal magnetic field is largely dipole-like) and for pure high-speed coronal-hole plasma (i.e. originated from open magnetic field lines), we find a deviation from the Parker model:
\begin{eqnarray}
|B_r| = (3.5\pm0.2) (R/R_0)^{-(1.81\pm 0.08)}\ \text{nT} \\
|B_t| = (2.9\pm0.2) (R/R_0)^{-(1.21\pm 0.09)}\ \text{nT}  
\end{eqnarray}
In particular, $B_r$ decreases more slowly than expected, while $B_t$ decreases faster (not shown here). It is worth remembering that Alfv\'enic fluctuations dominate these high-speed intervals that, due to their large amplitude and not linearly polarised characteristics, tend to preferentially reduce the radial fluctuations with respect to the background state.  
As a consequence, the radial behaviour of $B_r$ could be due to a combination of a mean part in accordance with a Parker spiral behaviour and a fluctuating part which a different scaling. 

The `failure' of the Parker prediction has been already reported in previous studies of in situ measurements from different missions~\cite[e.g.][]{loc09,kha12}. In particular, several works were dedicated to studying the slower decrease of $B_r$, implying to an excess of the magnetic flux, and many explanations, from kinematic effects to wrong averaging methods, were suggested~\citep{loc09,smi11,loc13}.
Recently,~\citet{kha12} found, by using multi-spacecraft hourly averaged data of magnetic field between $0.29$~AU and $5$~AU, that $|B_r| \propto R^{-1.666}$ and $|B_t| \propto R^{-1.096}$, showing a stronger deviation for $B_r$ than in our analysis. However,~\citet{kha12} selected the intervals in accordance with the radial distance; therefore, the data include both slow and fast wind and interaction regions. Their results suggest that turbulent processes in the inner heliosphere may significantly influence the expansion of the magnetic field, an idea already proposed by different authors~\citep[e.g.][]{rag06}. In particular, they propose a quasi-continuous magnetic reconnection, occurring at the large-scale heliospheric current sheet as well as at small-scale current sheets during the solar-wind expansion, as a key process responsible for breaking the expected magnetic field radial dependence law.

\subsection{Temperature}

Panel (c) of Fig.~\ref{fig:evolution} displays the radial dependence of the proton core temperature, $T_p$, where the fit (black-solid line) gives 
\begin{eqnarray}
T_p=(1.9\pm0.1) \cdot 10^5 (R/R_0)^{-(0.9\pm 0.1)}\ \text{K}
\end{eqnarray} 
The proton temperature decreases more slowly with respect to the adiabatic prediction, requiring an additional heat source. By considering a steady state, spherically expanding solar wind with a radially constant speed, a number density profile varying as $R^{-2}$ and a negligible proton heat flux, we find the following radial behaviour for the net volumetric heating rate~\citep{bre10}: $Q_p \propto R^{-(3.9\pm0.2)}$, in agreement with previous studies~\citep[e.g][]{hel11}. However, we observe that the profile of $T_p$ for unperturbed coronal-hole plasma decreases faster than in~\citet{hel11}, where $T_p \propto R^{-0.74}$ (black-dashed line); the inclusion of interaction regions could represent another source of heating. However, in the present study we consider only the proton core, while \citet{hel11} include both core and beam. The contribution of parallel and perpendicular temperature (not shown here) is 
\begin{eqnarray}
T_\parallel=(1.2\pm0.1) \cdot 10^5 (R/R_0)^{-(0.5\pm 0.1)}\ \text{K} \\
T_\perp=(2.3\pm0.1) \cdot 10^5 (R/R_0)^{-(1.0\pm 0.1)}\ \text{K}
\end{eqnarray} 
while~\citet{hel11} find $T_\parallel \propto R^{-0.54}$ and $T_\perp \propto R^{-0.83}$, respectively. The evolution of $T_\parallel$ is consistent but that of $T_\perp$ is not. Again, the slower decrease observed by~\citet{hel11} could be due to the presence of interaction regions in their data. 

It has been known for decades~\citep{mar82b,bou10} that in fast wind the proton core undergoes strong local perpendicular heating, with the temperature anisotropy, $T_\perp/T_{\parallel}$ reaching values of up to 3 at $0.3$~AU. $T_\perp/T_{\parallel}$ decreases with distance as shown in panel (d) of Fig.~\ref{fig:evolution} where the fit gives 
\begin{eqnarray}
T_\perp/T_{\parallel}=(2.1\pm0.1) (R/R_0)^{-(0.41\pm 0.08)} 
\end{eqnarray} 
still remaining larger than 1 at $1$~AU. This departure from an isotropic distribution function is a possible source of free energy for the proton cyclotron~\citep[e.g.][]{gary94} and mirror mode~\citep[e.g.][]{pok04} instabilities. 

\citet{gary01}, by using ACE measurements in the fast wind of protons (core plus beam), suggested that the proton cyclotron instability plays an important role in shaping the particle distribution function. ~\citet{mar04} have shown that the proton core does not appear constrained by the proton cyclotron instability but exhibits an anticorrelation between $T_\perp/T_{\parallel}$ and the proton-core parallel beta, $\beta_{\parallel}$, as 
\begin{eqnarray}
\label{eq:marsch}
T_\perp/T_{\parallel} \simeq \frac{a}{\beta_\parallel^b}  
\end{eqnarray} 
with $a \simeq 1.16$ and $b \simeq 0.55$. Moreover,~\citet{mat07} have shown that the correlation above is an evolution path followed during the expansion until 1~AU. Finally,~\citet{hel06} found that the constraint on $T_\perp/T_{\parallel} >1$ was better described by the mirror instability, where the instability threshold has been derived by calculating marginal stability (i.e. $\gamma = 10^{-3} \omega_{cp}$, where $\omega_{cp}$ is the proton cyclotron frequency) for $\beta_\parallel$ as
\begin{eqnarray}
\label{eq:instab}
T_\perp/T_{\parallel} = 1 + \frac{a}{(\beta_\parallel-\beta_0)^b}  
\end{eqnarray}  

Based on the $\beta_\parallel$ values in the stream A listed in Table~\ref{tab:intervals}, we can estimate the typical threshold values of the temperature anisotropy for the cyclotron [Eq.~(\ref{eq:instab}) with $(a,b,\beta_0)=(0.43, 0.42, -0.0004)$] and mirror mode [Eq.~(\ref{eq:instab}) with $(a,b,\beta_0)=(0.77, 0.76, -0.016)$] instabilities, and for the empirical relation [Eq.~(\ref{eq:marsch})] described in~\citet{mar04}. The results are shown in panel (d) of Fig.~\ref{fig:evolution}. In agreement with~\citet{hel06}, we found that the measured anisotropies are constrained by the mirror mode instability (pink-dot-dashed line), but the best agreement is found by using the anticorrelation proposed by~\citet{mar04} (black-dotted line), which follows closely the fit of $T_\perp/T_{\parallel}$. Finally, the values of temperature anisotropy do not seem to be limited by the proton cyclotron instability threshold (blue-dashed line). However, as pointed out by~\citet{ise13}, this result could be due to the estimation of the anisotropy threshold under the assumption that the proton distribution function is a bi-Maxwellian. In fact, resonant interactions between ion cyclotron waves and collisionless protons never yields bi-Maxwellian distributions.   

The presence of perpendicular heating can be also confirmed by the direct analysis of the radial evolution of the adiabatic invariants (not shown here). In particular, for the proton magnetic moment, we find  
\begin{eqnarray}
T_{\perp}/B = \mu_p \propto (R/R_0)^{(0.6\pm 0.1)}
\end{eqnarray}  
meaning that a significant trend for the first invariant to increase with increasing heliocentric distance is observed in pure fast wind, in agreement with the least squares fit index ($\sim 0.6$) found by~\citet{mar83} in the range of velocity $[700,800]$~km/s. In contrast, the fit for the second invariant gives
 \begin{eqnarray}
T_{\parallel}(B/n)^2 \propto (R/R_0)^{(0.2\pm 0.3)} 
\end{eqnarray}   
where, given the large error on the index, probably due to the bigger uncertainty on the parallel temperature, does not allow us to make conclusions about its conservation. Indeed,~\citet{mar83} found that, for the data corresponding to fast streams, the curve was almost flat. Finally, we can consider another invariant, $T_{\parallel} (T_\perp/n)^2$, which is independent of the three-dimensional structure of the interplanetary magnetic field~\citep{mar83}, under the assumption that energy sources and sink terms can be discarded, and we obtain
\begin{eqnarray}
T_{\parallel} (T_\perp/n)^2 \propto (R/R_0)^{(1.4\pm 0.4)} 
\end{eqnarray}    
in agreement with the behaviour observed by~\citet{mar83} in the velocity range $[700,800]$~km/s. Therefore, since at least one of the double-adiabatic invariants are observed to be broken, this indicates the action of dissipation or collisions.

\subsection{Pressure}

Another important property to characterise the solar wind is represented by the pressure. In particular, we can consider separately the contribution of the proton kinetic and magnetic pressures, whose radial dependences (not shown here) are
\begin{eqnarray}
P_k = (0.0065 \pm 0.0005) (R/R_0)^{-(2.9\pm 0.1)}\ \text{nPa} \\
P_B = (0.0131 \pm 0.0009) (R/R_0)^{-(3.2\pm 0.1)}\ \text{nPa}
\end{eqnarray}  
The magnetic pressure decreases faster with respect to the kinetic pressure in the expansion of the solar wind. This is in agreement with the observation of the proton plasma beta that increases as the radial distance increases 
\begin{eqnarray}
\beta_p = P_{k}/P_B = (0.55\pm0.04) (R/R_0)^{(0.4\pm 0.1)}
\end{eqnarray}     
even if the relative error on the exponent is about 25\%. However, the same behaviour is observed for the parallel proton plasma beta
 \begin{eqnarray}
\beta_{\parallel} = (0.37\pm0.03) (R/R_0)^{(0.8\pm 0.1)}
\end{eqnarray}
Here the relative error is less than 15\%.  

Finally, the values of the constant polytropic index in pure high-speed streams are $\gamma =1.47 \pm 0.22$ when the only contribution of the proton kinetic pressure is considered (see Eq.~6 in~\citet{tot95}), and $\gamma^{*} = 1.57 \pm 0.21$ when we consider the sum of proton kinetic and magnetic pressures (see Eq.~8 in~\citet{tot95}). Our results are in agreement with the values found by~\citet{tot95} for the speed range $[700,800]$~km/s.

\subsection{Collisional frequency}

\begin{figure}
	\includegraphics[width=\columnwidth]{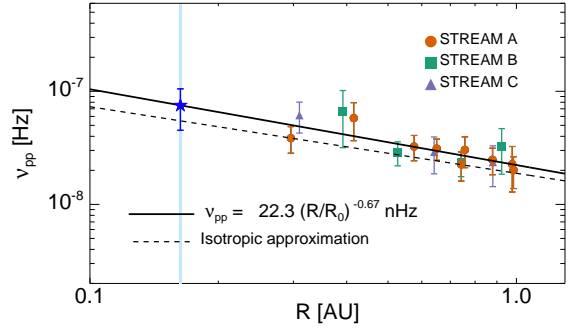}
    \caption{Radial dependence of the proton-proton collisional frequency. The legend is the same as in 
    				Fig.~\ref{fig:evolution}.}
    \label{fig:collisions}
\end{figure}

Although fast solar wind can be considered to a first approximation as a collisionless plasma, the observed temperature anisotropy cannot be described through a purely collisionless expansion, meaning that wave-particle interaction or heat conduction could play a role. However, there are observational evidences that collisional processes play an important role in the dynamics of the solar wind~\citep[e.g.][]{maruca13}. By using local plasma properties at the point of the observation it is possible to estimate the collisional frequency experienced by the interplanetary plasma, $\nu=1/\tau$, being $\tau$ the collisional time scale. 

Here, we calculate the collisional frequency of the proton core population, $\nu_{_{pp}}$, for a single bi-Maxwellian.  Following~\citet{hel16}, the isotropisation frequency is defined through the collisional evolution equations for parallel and perpendicular temperatures. The functional form of $\nu_{_{pp}}$, for relaxation of the perpendicular anisotropy ($T_\perp>T_\parallel$) in a gyrotropic ionised gas by Coulomb self-collisions, is~\citep{kog61}
\begin{eqnarray}
	\label{eq:coll}
            \nu_{_{pp}} = \frac{2\sqrt{\pi}e^{4}n_p\lambda}{m_p^{1/2}(k_B T_\parallel)^{3/2}}A^{-2} 
            \left[-3+(A+3)\frac{\tan^{-1}\sqrt{A}}{\sqrt{A}}\right]
\end{eqnarray}
where $A=T_\perp/T_\parallel-1$ and $\lambda$ is the Coulomb logarithm~\citep{chh16}.  
The radial evolution of $\nu_{_{pp}}$ is shown in Fig.~\ref{fig:collisions}, where the fit (black-solid line) gives
\begin{eqnarray}
	\nu_{_{pp}} = (22.3\pm2.2) \cdot 10^{-9} (R/R_0)^{-(0.7 \pm 0.2)}\ \text{s$^{-1}$}
\end{eqnarray}
The behaviour is monotonically decreasing with distance, although we observe a small bump between $0.3$ and $0.4$~AU (corresponding to both an increase in density and a depression in temperature). 
For reference, we plot also the radial evolution of $\nu_{_{pp}}$ but in a isotropic approximation (we assume that the isotropic temperature is equal to $T_p$), defined as $\bar{\nu}_{_{pp}}= 1.9 \cdot 10^{-8} n_p\lambda T_p^{-3/2 }$~s$^{-1}$, whose fit (black-dashed line) gives
\begin{eqnarray}
	\bar{\nu}_{_{pp}} = (18.8\pm1.9) \cdot 10^{-9} (R/R_0)^{-(0.6 \pm 0.2)}\ \text{s$^{-1}$}
\end{eqnarray}
The behaviour is the same but $\nu_{_{pp}}$ decreases faster than $\bar{\nu}_{_{pp}}$. Moreover, the presence of anisotropy produces locally a larger value for the collisional frequency, as suggested also by numerical experiments where an increase of plasma collisionality due to velocity space deformations of the particle velocity distribution functions is observed~\citep{pez16}.

Finally, if we extrapolate the value of $\tau$ from $\bar{\nu}_{_{pp}}$ to $0.1$~AU, we find $\sim 1.4 \cdot 10^7$~s, larger than the value found in~\citet[]{chh16} (see Fig. 1), where a three-dimensional magnetohydrodynamic simulation of the global heliosphere has been used to improve the estimation of the collisional time during the solar wind expansion by means of the self-collisional time defined in~\citet[]{spi56}. This suggests that their model slightly overestimates the collisionality with respect to observations.

\section{Discussions and Conclusions}

We have presented a detailed analysis of the radial evolution of homogeneous and unperturbed coronal-hole plasma from 0.3~AU ($\sim 60 R_s$) to the Earth, by using re-processed proton-core HELIOS data. 

\begin{table}
	\centering
	\begin{threeparttable}
	\caption{Comparison of the radial evolution index for several quantities and 
			polytropic indices in fast solar wind. No information about the error is 
			given in some of the previous analyses. The data for the unperturbed
			coronal-hole plasma are from the stream A listed in Table~\ref{tab:intervals}}
	\label{tab:comparison}
	\begin{tabular}{cccc}
		\hline
		\multicolumn{4}{ c }{Radial evolution index} \\
		\hline
		            & Unperturbed  & Fast wind  & Fast wind   \\
		            & coronal-hole & $V > 600$~km/s & speed range \\
		            & plasma & \citet{hel11} & [700, 800]~km/s\tnote{a} \\ 
		\hline	
		$n_p$ & $-1.96\pm 0.07$ & -1.8 & $-1.88\pm0.23$ \\
		$B$  & $-1.59\pm 0.06$ & -1.6 & $-1.45\pm0.20$  \\
		$T_p$ & $-0.93 \pm 0.10$ & -0.74 & $-0.86\pm0.28$ \\
		$T_{\parallel}$ & $-0.53\pm 0.13$ & -0.54 & -0.69 \\
		$T_{\perp}$ & $-1.01 \pm 0.10$ &  -0.83 & -1.17\\
		$T_{\perp}/B$ & $0.59 \pm 0.10$ & - & 0.6\\
		$T_{\parallel}(B/n)^2$ & $0.21 \pm 0.26$ & - & $\sim$ const \\
		$T_{\parallel} (T_\perp/n)^2$ & $1.36 \pm 0.35$ & - & 0.6 \\
		$\beta_p$ & $0.36\pm 0.13$ & - & $0.61\pm0.18$ \\
		$\beta_{\parallel}$ & $0.81\pm0.14$ & $1.91\pm2.55$\tnote{b} & - \\ 		\hline
		$\gamma$  & $1.47\pm0.22$ & - & $1.46\pm0.16$\\
		$\gamma^{*}$   & $1.57\pm0.21$ & - & $1.51\pm0.20$ \\
		\hline		
	\end{tabular}
	\begin{tablenotes}\footnotesize
		\item [a] Different references have been used for comparison. In particular, the parameters with the 
				error are from~\citep[]{tot95}, while the others are from~\citep[]{mar82b} 
				and~\citep[]{mar83}.
		\item [b] This value is from~\citep[]{marric84}.
	\end{tablenotes}
	\end{threeparttable}
\end{table}

A correct classification of the solar wind is crucial in order to understand and interpret the observations. Most of the studies on the radial evolution of the solar wind in different regimes have been based on the particle velocity classification. However, this choice can be too simplistic, since no exact speed thresholds exist to distinguish between fast and slow streams. 

Several recent papers have considered the classification of solar wind based on its origin~\citep[][and references therein]{cam18}. Here, we concentrated on unperturbed plasma from coronal holes. The selection method, based on the density, velocity, magnetic field and proton specific entropy~\citep{bor16}, allow us to avoid the interaction regions, which represent an external source of heating and compression. In fact, if the choice of fast intervals, e.g. for stream A, is done only by inspection of the plasma speed (i.e. the interval is considered between the highest value of the velocity after the abrupt increase and value before the systematically decrease), we find a change in the radial evolution of the streams that confirms the presence of compressed regions, e.g. the density decreases as $R^{-1.92}$ instead of $R^{-1.98}$ when we fit all the points. Moreover, we have tried to apply the threshold of $V > 600$~km/s~\citep{hel11} on the two-years of HELIOS data considered in the present analysis and we found slight differences with respect to the unperturbed intervals. However, when the threshold is applied to many years of solar wind data~\citep{sta18}, the results are almost comparable with~\citet{hel11} (even if some differences can be due to the presence or not of the proton beam in the data). Therefore, although the threshold in velocity could be a good approximation for the choice of fast stream in case of short intervals, it might become limited and misleading in case of large statistical datasets.     
 
The other idea of the present analysis is to follow well-defined streams during several solar rotations and at different distances from the Sun, with the assumption that the coronal holes, where these streams originated, were stable in time. Recently,~\citet{hei18} have shown that the properties of high-speed streams (especially the solar wind proton bulk velocity) measured at 1~AU are affected by the evolution of the coronal holes observed in the solar atmosphere. However, their analysis is based on the observation of a well-observed, long-lived and low-latitude coronal hole in the year 2012, corresponding to maximum of solar activity. In our analysis, we are considering a period of solar minimum, where the behaviour of the magnetic field on the Sun should be more regular and stable. Unfortunately, no data from the solar atmosphere are available in order to study the evolution of coronal holes. However, it is possible to use data from IMP8 satellite as a reference for the high-speed stream properties at a fixed distance (i.e. 1~AU) complementary to the HELIOS satellites, to have an idea of the overall evolution of the high-speed streams. The comparisons (not shown here) confirm that the coronal holes are almost stable, since no strong change in the properties of the streams are observed. Therefore, we can conclude that in our intervals we are only sensitive to the radial, and not temporal, evolution of the unperturbed coronal-hole plasma.  

\begin{table}
	\centering
	\caption{Summary of the fitting for the radial profile of different quantities,
			$f=f_0 (R/R_0)^{\alpha}$ (where $R_0=1$~AU),  
			for unperturbed high-speed coronal-hole plasma.
			All the intervals listed in Table~\ref{tab:intervals} are used together 
			independently of their origin.}
	\label{tab:final}
	\begin{tabular}{ccc}
		\hline
		\multicolumn{3}{ c }{Unperturbed coronal-hole plasma evolution}  \\
		\hline
		\cline{2-2} \cline{3-3}
                   & {$f_0$} & {$\alpha$} \\
		\cline{2-2} \cline{3-3}
		\hline
		$n_p\ [\text{cm}^{-3}]$ & $2.37\pm0.08$ & $-2.02 \pm 0.05$	\\
		$n_p V_r\ [10^3 \ \text{cm$^{-3}$ km/s}]$ & $1.53\pm0.05$ & $-1.99\pm 0.05$ \\
		$B\ [\text{nT}]$ & $5.6\pm0.1$ & $-1.63\pm 0.03$ \\
		$|B_r|\ [\text{nT}]$ & $3.5\pm0.1$ & $-1.84\pm 0.05$ \\
		$|B_t|\ [\text{nT}]$ & $2.9\pm0.1$ & $-1.29\pm 0.06$ \\
		$T_p\ [10^5\text{K}]$ & $1.96\pm0.09$ & $-0.90\pm 0.08$ \\
		$T_\parallel\ [10^5 \ \text{K}]$ & $1.29\pm0.07$ & $-0.48\pm 0.09$ \\
		$T_\perp\ [10^5 \ \text{K}]$ & $2.3\pm0.1$ & $-0.99\pm 0.08$ \\
		$T_\perp/T_\parallel$ & $2.02\pm0.07$ & $-0.43\pm 0.06$ \\ 
		$T_{\perp}/B\ [10^4\ \text{K/nT}]$ & $4.2\pm0.2$ & $0.65\pm 0.08$ \\
		$T_{\parallel}(B/n)^2\ [10^5 \ \text{K nT$^2$ cm$^{6}$}]$ & $7.6\pm0.7$ & $0.3\pm 0.2$ \\
		$T_{\parallel} (T_\perp/n)^2\ [10^{15} \ \text{K$^3$ cm$^{6}$}]$ & $1.4\pm0.2$ & $1.6\pm 0.3$ \\
		$\nu_{_{pp}}\ [\text{nHz}]$ & $22\pm2$ & $-0.8 \pm 0.1$ \\
		$ S_p\ [\text{eV cm$^2$}]$ & $9.6\pm0.5$ & $0.45\pm 0.09$ \\
		$P_{k}\ [10^{-11}\ \text{Pa}]$ & $0.66\pm0.04$ & $-2.89\pm 0.09$ \\
		$P_{B}\ [10^{-11}\ \text{Pa}]$ & $1.30\pm0.06$ & $-3.23\pm 0.07$ \\
		$\beta_p$ & $0.56\pm0.03$ & $0.40\pm 0.08$ \\
		$\beta_{\parallel}$ & $0.40\pm0.03$ & $0.9\pm 0.1$ \\
		\hline		
	\end{tabular}
\end{table}

The results of our analysis on the radial evolution of 9 unperturbed high-speed intervals from a specific coronal hole (i.e. stream A in Table~\ref{tab:intervals}) is reported in Table~\ref{tab:comparison}, where a comparison with previous analyses in fast wind is given. We found that (i) the proton density decreases as expected for a stationary radially expanding plasma, while previous analyses have shown a slower decrease; (ii) the evolution of the magnetic field is similar to previous analyses. However, by looking in detail at the evolution of the magnetic components, we found a deviation from the Parker prediction, with the radial component decreasing more slowly and the tangential component decreasing faster than expected. (iii) We observed a perpendicular heating in the proton core population, with a violation of the double-adiabatic invariants and a corresponding increase of entropy. The same conclusions can be obtained from previous studies, although different indices are recovered. (iv) We studied the perpendicular heating through the evolution of the proton core temperature anisotropy and we found that it is constrained in the radial expansion of the fast solar wind by the mirror mode instability, consistent with~\citet[]{hel06}. However, the best agreement is found with the empirical relation proposed by~\citet[]{mar04}. (v) Proton kinetic and magnetic pressures fall in a way that the proton plasma beta increases with distance. Moreover, we found characteristic constant polytropic indices that are in agreement with previous studies of high-speed streams. (vi) The behaviour of the collisional frequency for a bi-Maxwellian distribution is to decrease as the plasma moves away from the Sun. Moreover, comparing $\nu_{_{pp}}$ with its isotropic approximation we found a faster decrease and larger local values. Finally, we observe that our conclusions are almost independent, in the limit of the error, of the specific coronal-hole source. In this respect, in Table~\ref{tab:final} we summarise the radial dependence (power-law index and expected value at 1~AU) for all the parameters evaluated in our study. 

\begin{table}
	\centering
	\caption{Prediction for fast solar wind during the first three 
			perihelia of Parker Solar Probe.}
	\label{tab:prediction}
	\begin{tabular}{ccc}
		\hline
		\multicolumn{3}{ c }{Unperturbed coronal-hole plasma at $35 R_s$}  \\
		\hline
		\hline
		$B$ & $|B_r|$ & $|B_t|$ \\
		$(108 \pm 9)$ nT & $(99 \pm 11)$ nT & $(30 \pm 4)$ nT \\
		\hline
		$T_p$ & $T_\perp$ & $T_\parallel$ \\
		$(10 \pm 2) \ 10^5$ K & $(14 \pm 3) \ 10^5$ K & $(3.1 \pm 0.7) \ 10^5$ K \\
		\hline 
		$n_p$ & $\nu_{_{pp}}$ & $\beta_p$ \\
		$(92 \pm 12)$ cm$^{-3}$ & $(9 \pm 3) \ 10^{-8}$ Hz & $(0.31 \pm 0.04)$ nHz \\
		\hline
		$P_{k}$ & $P_B$ & $\beta_{\parallel}$ \\
		$(1.3 \pm 0.3)$ nPa & $(4.6 \pm 0.8)$ nPa & $(0.08 \pm 0.01)$\\
		\hline
		\hline		
	\end{tabular}
\end{table}

Observations of fast solar wind from the three orbits of the Ulysses mission have shown that the solar wind energy flux and the particle flux are regulated by the amount of magnetic flux that opens into the heliosphere~\citep{sch08}. Parker Solar Probe and Solar Orbiter will give an important contribution to look for secular changes. Our work fits perfectly in the context of the next solar missions. The knowledge of how plasma from coronal holes evolves between $0.3$ and $1$~AU allows us to predict the high-speed solar-wind environment much closer to the Sun, which Parker Solar Probe will explore in the near future. In particular, during the first three perihelia (until September 2019), Parker Solar Probe will take measurements of the solar wind plasma at heliocentric distance of $\simeq 0.163$~AU ($\sim 35 R_s$), before approaching the Alfv\'en critical point. In Table~\ref{tab:prediction} we estimate the characteristic mean values of fast solar-wind plasma at the distance of $35 R_s$ by using the radial trends in~Table~\ref{tab:final}. However, only Solar Orbiter could provide an insight into the problem of the origin for different streams, by means of the combination of in situ and remote sensing observations.

\section*{Acknowledgements}

Work by D. Perrone and T. S. Horbury was supported by STFC grant ST/K001051/1; D. Stansby was supported by a studentship under STFC grant ST/N504336/1; and L. Matteini was supported by the Programme National PNST of CNRS/INSU co-funded by CNES.







\bsp	
\label{lastpage}
\end{document}